\documentclass[a4paper,11pt]{article}
\usepackage{pos}
\usepackage{sidecap}

\title{VERITAS Observations of the Galactic Center Region at Multi-TeV Gamma-Ray Energies}
 \ShortTitle{VERITAS Observations of the Galactic Center Region at Multi-TeV Gamma-Ray Energies}

\author*[a]{James L. Ryan}
\forColl{VERITAS}

\affiliation[a]{University of California, Los Angeles}
\affiliation[b]{\url{veritas.sao.arizona.edu}}

\emailAdd{jryan@astro.ucla.edu}

\abstract{
The Galactic Center region hosts a variety of powerful astronomical sources and rare astrophysical processes that emit a large flux of non-thermal radiation. We present the analysis of the very-high-energy gamma-ray emission above 2 TeV of the region around the Galactic Center known as the Central Molecular Zone using 125 hours of data taken with the VERITAS imaging-atmospheric Cherenkov telescope between 2010 and 2018. This analysis employs new shower reconstruction algorithms and instrument response functions optimized for data taken at large zenith angles such as the Galactic Center sources. We report positions and spectra for point sources VER J1745-290, G0.9+0.1, and HESS J1746-285, along with a light curve for VER J1745-290, the brightest source in the region consistent with the position of the supermassive black hole Sagittarius A*. We also measure the spectrum of the diffuse emission from the Galactic Center ridge region, which has been claimed as evidence of a Galactic PeVatron.
}

\FullConference{37$^{\rm{th}}$ International Cosmic Ray Conference (ICRC 2021)\\
		July 12th -- 23rd, 2021\\
		Online -- Berlin, Germany}

\def\degr{^\circ}
\def\dflux{\text{ cm}^{-2}\text{ s}^{-1}\text{ TeV}^{-1}}

\begin{document}
\maketitle

\section{Introduction}
The region within a few degrees of the Galactic Center (GC) hosts many sources of astrophysical interest that may emit very-high-energy (VHE; $\gtrsim$100 GeV) $\gamma$-rays.
Such sources include the supermassive black hole Sagittarius A* (hereafter Sgr A*), supernova remnants (SNRs), pulsar wind nebulae (PWNe), and dense molecular clouds.
The VHE emission from these objects is crucial in understanding their emission mechanisms, and can serve as one of the few signatures of a source emitting PeV cosmic rays.

We analyze GC observations taken with the Very Energetic Radiation Imaging Telescope Array System (VERITAS), an imaging atmospheric Cherenkov telescope (IACT) sensitive to $\gamma$-rays in the energy range of 100 GeV to above 30 TeV.
Using new data and improved analysis techniques, we present an updated analysis of TeV sources in the GC region.
Full details of the analysis can be found in \citep{adams2021}.

\section{Observations}
Observations of the GC region were taken with VERITAS, located at the Fred Lawrence Whipple Observatory (FLWO) in southern Arizona (31$^\circ$ 40' N, 110$^\circ$ 57' W,  1.3km above sea level). 
Due to VERITAS' location, all GC observations are taken at large zenith angles (LZA), $\gtrsim 59^\circ$.
Data are taken at $0.5^\circ$ and $0.7^\circ$ offsets from the position of Sgr A*, in right ascension or declination. 
155 hr of exposure time was accumulated between 2010 April and 2018 June, of which 125 hr of data remain after quality cuts and dead time correction.
This dataset contains an additional 40 hr more than the previous VERITAS analysis of the GC region \citep{archer2016}.

\section{Analysis}
Whereas typical VERITAS analyses concern point sources observed at small zenith angles, the present analysis of the Galactic Center consists only of LZA observations, and contains the highly extended source of the diffuse ridge emission.
Observing at LZA, compared with smaller zenith angles, has the established effects of increasing an IACT's effective area at higher energies while raising the minimum energy threshold of detectable $\gamma$-rays. 
However, we find that LZA data also present several challenges to the standard VERITAS analysis pipeline: shower direction reconstruction worsens, a bias arises in energy reconstruction, and a zenith-dependent component in the acceptance map (the spatially-dependent expected number of background events) becomes significant.
We introduce new techniques to the VERITAS analysis pipeline to address these issues.

The \textit{geometrical} method 
of determining shower direction, which uses the intersection of the major axes of the approximately elliptical shower images in each telescope, performs well at small zenith angles but worsens considerably at zenith angles greater than $40\degr$ \citep{archer2014}.
We instead use an algorithm based on the \textit{displacement} method, which estimates the distance between the shower direction and a single image centroid (the \textit{disp} parameter) using information from many simulated events.
Our algorithm estimates the \textit{disp} parameter using a boosted decision tree (BDT) algorithm trained on simulations, using as features image size, width, length, difference in photon arrival time, and the sum of the signals in all edge pixels (this correlates with the amount of Cherenkov light falling outside the camera).

Even with improved direction estimates from the BDT \textit{disp} method, event energy reconstruction at LZA still suffered from a $\sim$20\% systematic bias, determined from simulations.
We developed another BDT algorithm to estimate energy, using the same simulations and similar features.
Using both BDT algorithms, the energy bias is reduced to approximately zero above the energy threshold of $\sim$2 TeV,
and uncertainties become less that $20\%$, comparable to energy resolution at SZA ($15-25\%$).

The last LZA problem we correct for is the zenith dependence appearing in acceptance maps, which are assumed to be radially symmetric.
We fit the 2D maps of background gamma-ray like events (excluding regions around known TeV sources and bright stars) with radial and zenith-dependent components (both 4th-order polynomials).
The results of this fit determine the 2D acceptance map.

An additional optimization is made to the event-selection box cuts for LZA---based on Crab data taken at zenith angles similar to our GC dataset.
The LZA-optimized cuts give a 10\% increase in sensitivity versus standard cuts.
Instrument response functions for the effective telescope area are then generated using simulations analyzed with the methods just described, and stored in ROOT histograms.
The effective areas are used in the spectral flux calculations.
We calculate the binned differential energy spectra for each source, and also parametrize the spectra using one of three models: a power law (PL)
\begin{equation}
dN/dE=N_0 \left(\frac{E}{E_0}\right)^\Gamma
\end{equation}
an exponentially cut-off power law (ECPL)
\begin{equation}
dN/dE=N_0 \left(\frac{E}{E_0}\right)^\Gamma \mathrm{Exp}\left(-\frac{E}{E_\text{cut}}\right)
\end{equation}
or a smooth broken power law
\begin{equation}
dN/dE=N_0 \frac{\left(E/E_0\right)^{\Gamma_1}}{1+\left(E/E_\text{break}\right)^{\Gamma_1-\Gamma_2}}
\end{equation}
and report the best-fit parameters of the most parsimonious model that provides an adequate fit, based on $\chi^2$.
We validate our full methodology on our LZA Crab dataset, ensuring that the spectrum is in agreement with the spectrum resulting from a standard analysis at small zenith angles.


\section{Results}
\begin{figure}[t!]
\centering
\includegraphics[width=0.95\textwidth]{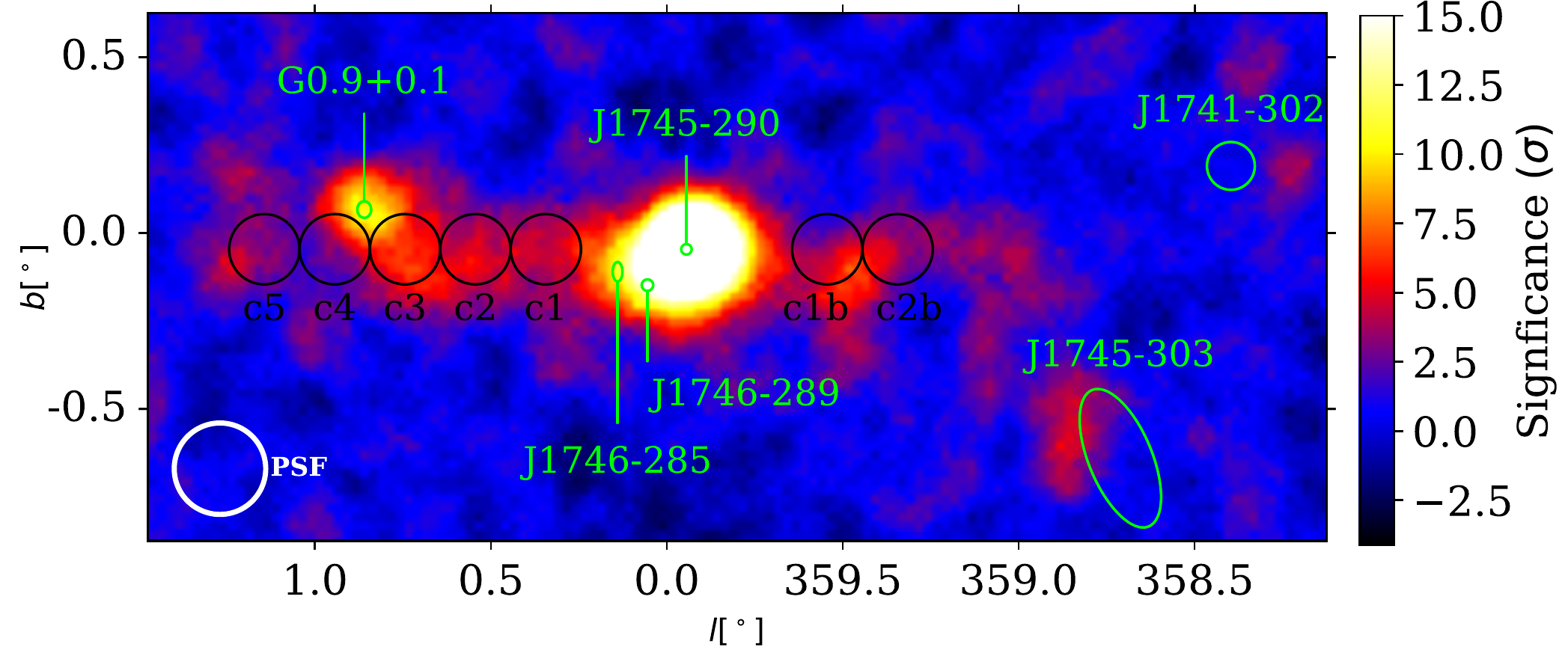}
  \caption{Map of the statistical significance for gamma-ray-like events above 2 TeV detected by VERITAS.
  Each pixel displays the significance of excess counts integrated over a $0.1\degr$ circular signal region.
  The VERITAS point spread function is shown in the bottom-left.
  Positions and 68\% confidence regions are shown for previously detected point sources (green ellipses),
  while the ellipses for J1741-302 and J1745-303 represent their spatial extents \citep{abdalla2018b,aharonian2006a}.
  The signal regions used in the diffuse ridge analysis are shown as black circles, and labeled as in \citep{abramowski2016}.
  \label{fig:gc_map}}
\end{figure}
We analyze our GC dataset using the new methods we have outlined.
A significance map of the inner $3\degr \times 1.5\degr$ of the GC region is shown in Figure \ref{fig:gc_map}.
The value in each pixel represents the significance of the excess counts in a $0.1\degr$ circular signal region centered on the pixel, with background counts estimated with the ring background and zenith-corrected acceptance map.

We find significant detections for J1745-290, G0.9+0.1, J1746-285, and the diffuse ridge emission, and measure their positions and spectra.
J1745-303 and J1741-302 are not significantly detected.

\subsection{J1745-290}
\begin{figure}[!t]
 \centering
  \includegraphics[width=0.9\textwidth]{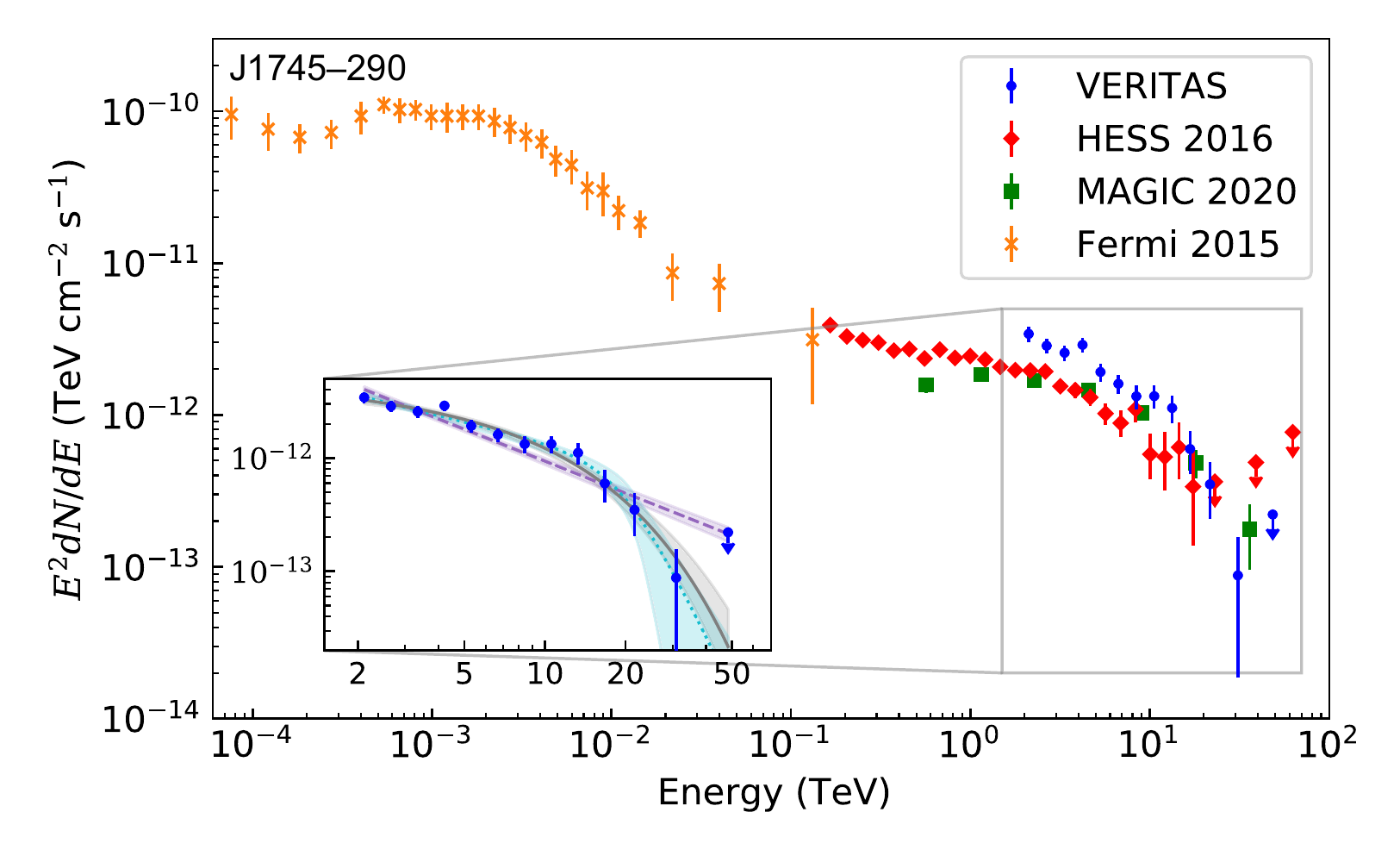}
  \caption{Differential energy spectrum of the central source J1745-290, coincident with the position of Sgr A*, as measured by VERITAS (blue), H.E.S.S. (red) \citep{abramowski2016}, MAGIC (green) \citep{acciari2020}, and \textit{Fermi}-LAT (orange) \citep{malyshev2015}.
  Error bars represent $1\sigma$ uncertainties in the flux.
  The downward arrows represent 95\% upper limits.
  In the inset, the best-fit exponentially cutoff power law (gray solid line), smooth broken power law (cyan dotted line), and power law (purple dashed line) are shown.
  Shaded regions represent the $1\sigma$ confidence band on the model fits.
\label{fig:j1745-290_spectrum} }
\end{figure}

We detect the bright central source J1745-290 with 37.5$\sigma$ significance, and measure a best-fit position of $(l,b)=(359.930^\circ,-0.047^\circ)$ with uncertainties in $l$ and $b$ of $0.018\degr$.
It is spatially coincident with Sgr A*, but remains without a definitive association as alternative potential TeV sources fall within its positional uncertainty (e.g. PWN G359.95-0.04).
The differential energy spectrum is shown in Figure \ref{fig:j1745-290_spectrum}.
Its spectrum can be described by an ECPL with $N_0=1.27^{+0.22}_{-0.23}\times 10^{-13}\dflux$, $\Gamma=-2.12^{+0.17}_{-0.22}$, $E_0=5.3$ TeV, and $E_\text{cut}=10.0^{+4.0}_{-2.0}$ TeV, in the energy range 2.5--40 TeV.
Our flux normalization is slightly higher than those of HESS or MAGIC (though within systematic uncertainties), due in part to their handling of flux contributions from nearby sources.

\begin{SCfigure}[.7][!t]
\centering
  \includegraphics[width=0.5\textwidth]{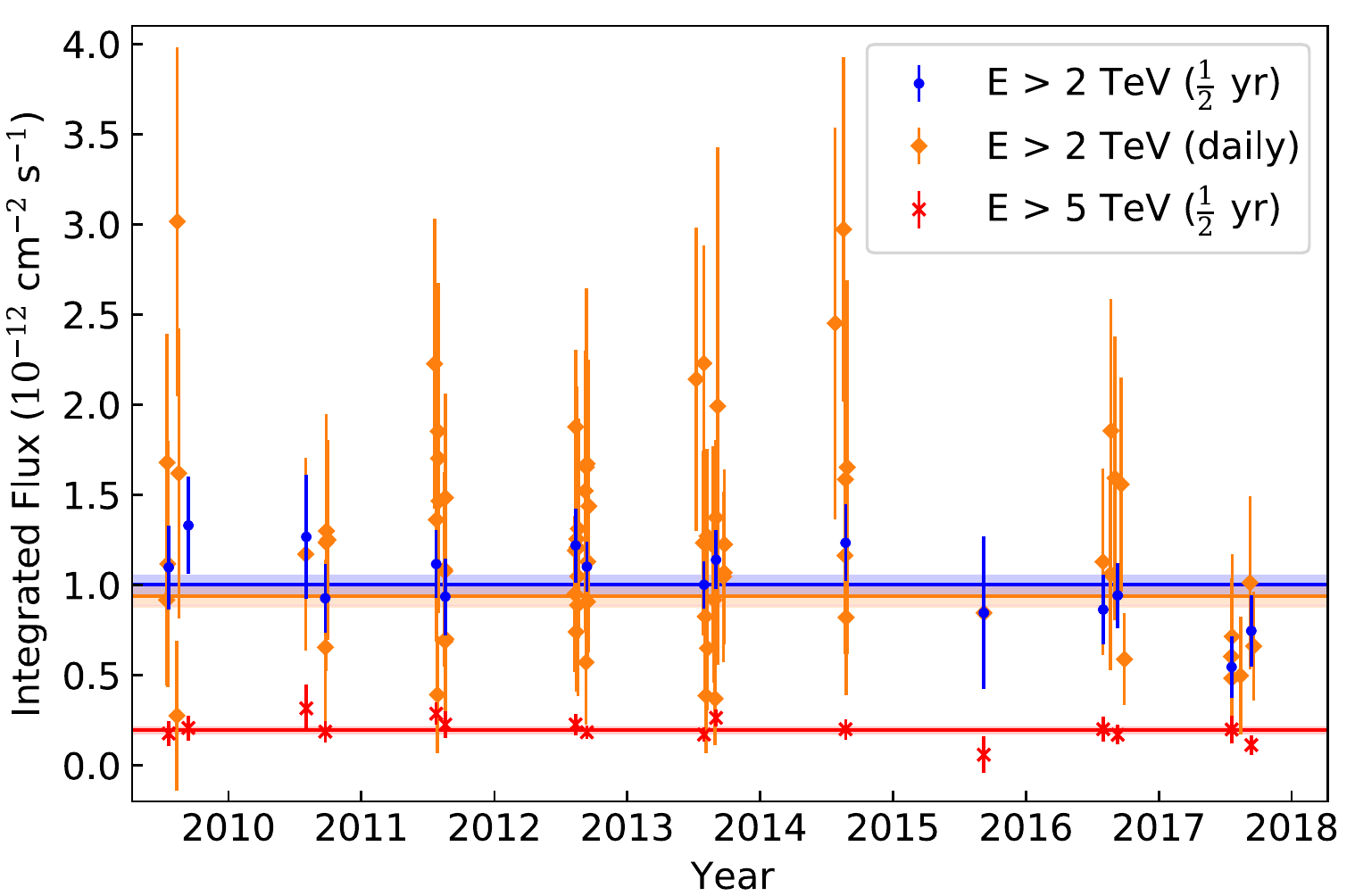}
  \caption{Semiannually (blue) and daily (orange) binned light curves of the integral flux above 2 TeV of J1745--290, showing flux versus time.
  Also shown is the semiannually-binned light curve above 5 TeV (red).
  The weighted means are shown as horizontal lines of the corresponding color, while the shaded regions indicate the 68\% confidence interval on the mean.}
  \label{fig:j1745-290_lightcurve}
\end{SCfigure}
The detection of any flux variability in J1745-290, such as characteristic timescales or variability correlated with other wavelengths, would improve our understanding of the source, so we produce light curves by performing our spectral analysis on time-interval subsets of the data.
Light curves for the integrated flux above 2 TeV, with both day and six-month binning, as well as the semiannually-binned light curve for integrated flux above 5 TeV, are shown in Figure \ref{fig:j1745-290_lightcurve}.
All light curves are found to be consistent with constant flux hypotheses.
A typical 95\% upper limit on the $>$2 TeV flux for a night with a 1 hr exposure is $\sim$$1.5\times10^{-12}\text{ cm}^{-2}\text{ s}^{-1}$.

Although interpretation of the spectrum is complicated by its uncertain source association, the VHE spectrum still can provide important constraints on any source (or sources) in this region, whether it be Sgr A*, pulsars, or dark matter.
If the diffuse ridge emission is produced by particles accelerated near the GC, it is also necessary to model both spectra consistently.

\subsection{Diffuse Ridge Emission} \label{sec:ridge}
\begin{SCfigure}[1][!b]
 \centering
  \includegraphics[width=0.5\textwidth]{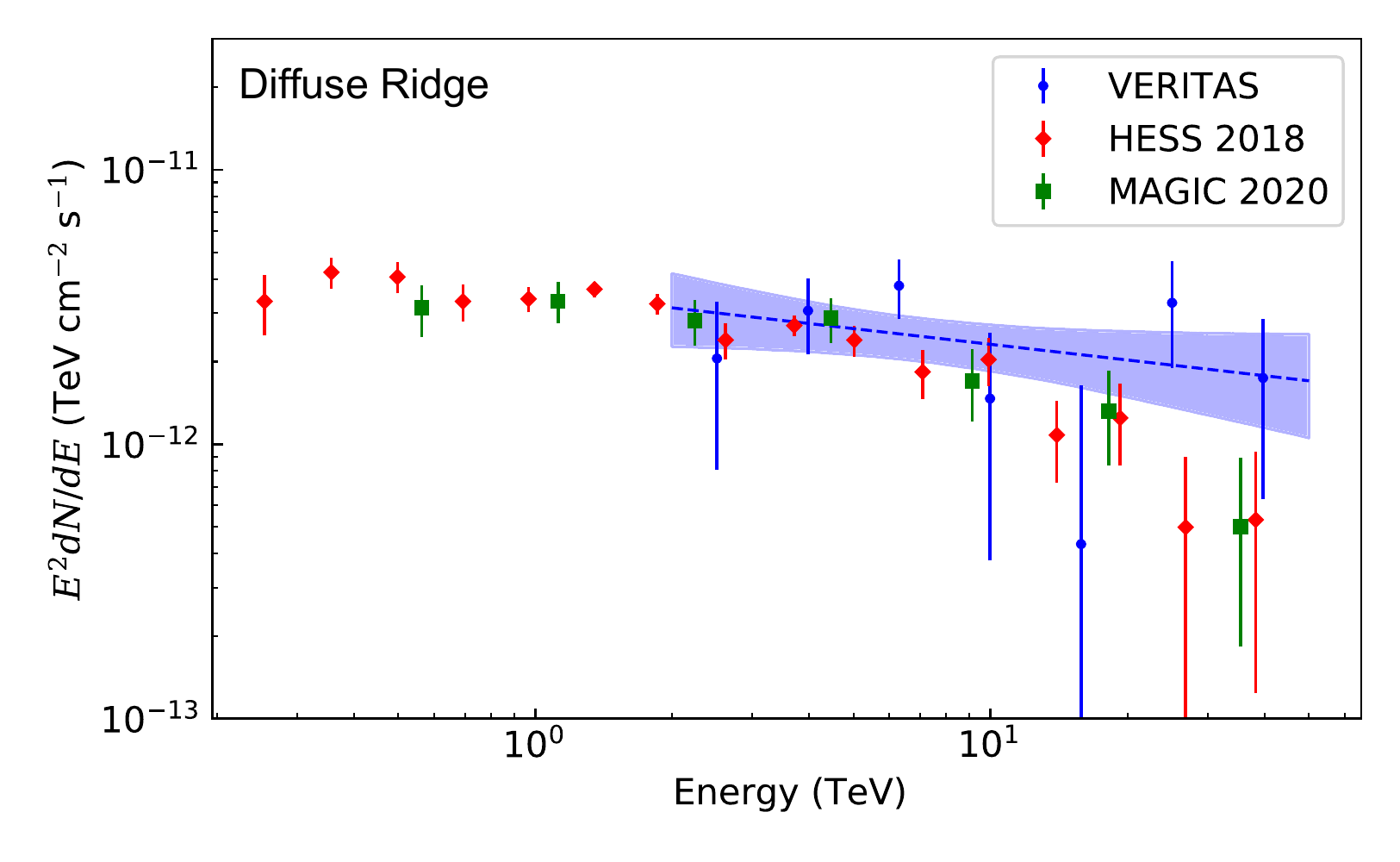}
  \caption
  {VHE $\gamma$-ray spectrum of the diffuse ridge emission measured by VERITAS (blue) from the combined circular regions along the GC ridge, with the normalization scaled to what would be measured in the signal region used by \citep{abdalla2018b}, as described in Section \ref{sec:ridge}. For comparison, the spectra measured by H.E.S.S. (red) \citep{abdalla2018b} and MAGIC (green) \citep{acciari2020} are shown.
  The blue dashed line is the best-fit power law with index $-2.19 \pm 0.20$
  The shaded region represents the $1\sigma$ confidence band on the model fit.
  \label{fig:diffuse_spectrum_ver_hess}}
\end{SCfigure}
TeV emission along the Galactic ridge close to the GC was first detected in 2006 \citep{aharonian2006b}, and was found to be spatially correlated with the molecular gas in the inner several hundred parsecs around the GC.
Standard methods for determining the number of events in signal and background regions for point sources, such as ring background or reflected region (``wobble'') methods, are unsuitable for the analysis of the diffuse ridge emission---having difficulty covering the full region of interest while avoiding the inclusion of signal in the background region(s).
We therefore combine the statistics of seven circular signal regions (the same as those used in \citep{abramowski2016}), and use non-overlapping reflected regions that lie away from the Galactic plane to estimate background counts.
The first VERITAS measurement of the diffuse ridge emission spectrum in shown in Figure \ref{fig:diffuse_spectrum_ver_hess}.
Its spectrum can be described by a power law with $N_0=(3.44\pm 0.62)\times10^{-14}\dflux$, $\Gamma=-2.19\pm0.20$, and $E_0=5.3$ TeV, in the energy range 2--20 TeV.
In the figure, we scale up our spectrum to visually compare it with other measurements that use larger signal regions.
The scale factor we use is the ratio of the integral of a model diffuse ridge flux map over the signal region used in \citep{abdalla2018b} to the integral over our signal region, using the velocity-integrated CS 1-0 map \citep{tsuboi1999} convolved with the VERITAS LZA point spread function as the model map.
The scale factor is found to be 2.7.

The diffuse ridge emission spectrum and spatial profile have been interpreted as evidence for a nearby accelerator of protons to PeV energies \citep{abramowski2016}.
The search for Galactic sources of PeV particles (in particular up to the ``knee'' at 3 PeV in the all-particle spectrum), or ``PeVatrons,'' is of great interest in cosmic ray physics. 
We fit the $\gamma$-ray spectrum resulting from a PL proton spectrum, using the parameterization in \citep{kelner2008}, and find a best-fit index of $\Gamma=-2.3$.
We then calculate the lower limit on the cutoff energy for an ECPL proton spectrum with index fixed to $-2.3$, and find a  95\% lower limit of $E_\text{cut}= 0.08$ PeV.

\subsection{G0.9+0.1} 
G0.9+0.1 is a ``composite SNR,'' composed of a SNR shell surrounding a PWN core.
It is young enough (order of kyr) 
that the PWN has not been influenced by a SNR reverse shock 
which, combined with its extensive multiwavelength coverage, makes it a common PWN to model. 

We detect G0.9+0.1 with 8$\sigma$ significance, and measure a best-fit position of $(l,b)=(0.857^\circ,0.069^\circ)$ with uncertainties in $l$ and $b$ of $0.033\degr$.
Its spectrum can be described by a power law with $N_0=1.51\dflux$, $\Gamma=-2.00\pm0.28$, and $E_0=5.3$ TeV, in the energy range 2--20 TeV.

\begin{SCfigure}[0.55][!b]
 \centering
  \includegraphics[width=0.6\textwidth]{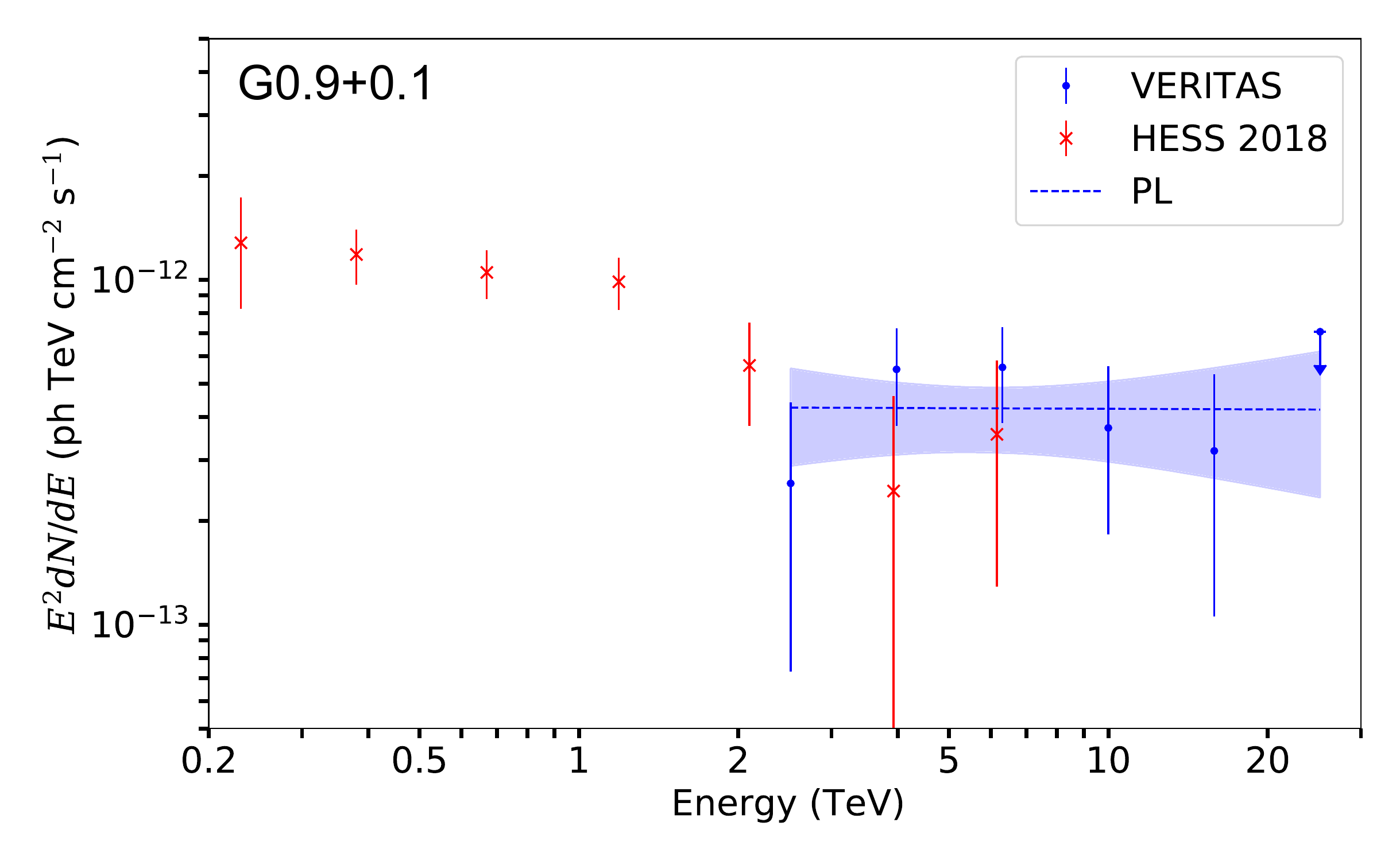}
  \caption{Differential energy spectrum of G0.9+0.1, as measured by VERITAS (blue), with the H.E.S.S. spectrum shown in red \citep{abdalla2018a}.
  Error bars represent $1\sigma$ uncertainties in the flux.
  The downward arrow represents a 95\% upper limit.
  The blue dashed line is the best-fit power law with index $-2.00 \pm 0.28$
  The shaded region represents the $1\sigma$ confidence band on the model fit.}
  \label{fig:g09_spectrum}
\end{SCfigure}

While spectral features such as bends or cutoffs would provide stronger model constraints, our spectrum does not significantly deviate from a simple power law, even taken together with the measurements by HESS \citep{abdalla2018a} and MAGIC \citep{acciari2020}.
In one model, a cutoff lying at 20-30 TeV is been predicted \citep{fiori2020}, just beyond the extent of our measured spectrum.
Understanding PWN remains especially important with respect to cosmic rays, due to their strong connection with SNR, as well as their expected contributions to the leptonic CR spectra.

\subsection{J1746-285} 
\begin{SCfigure}[1.0][!t] \centering
  \includegraphics[width=0.6\textwidth]
  {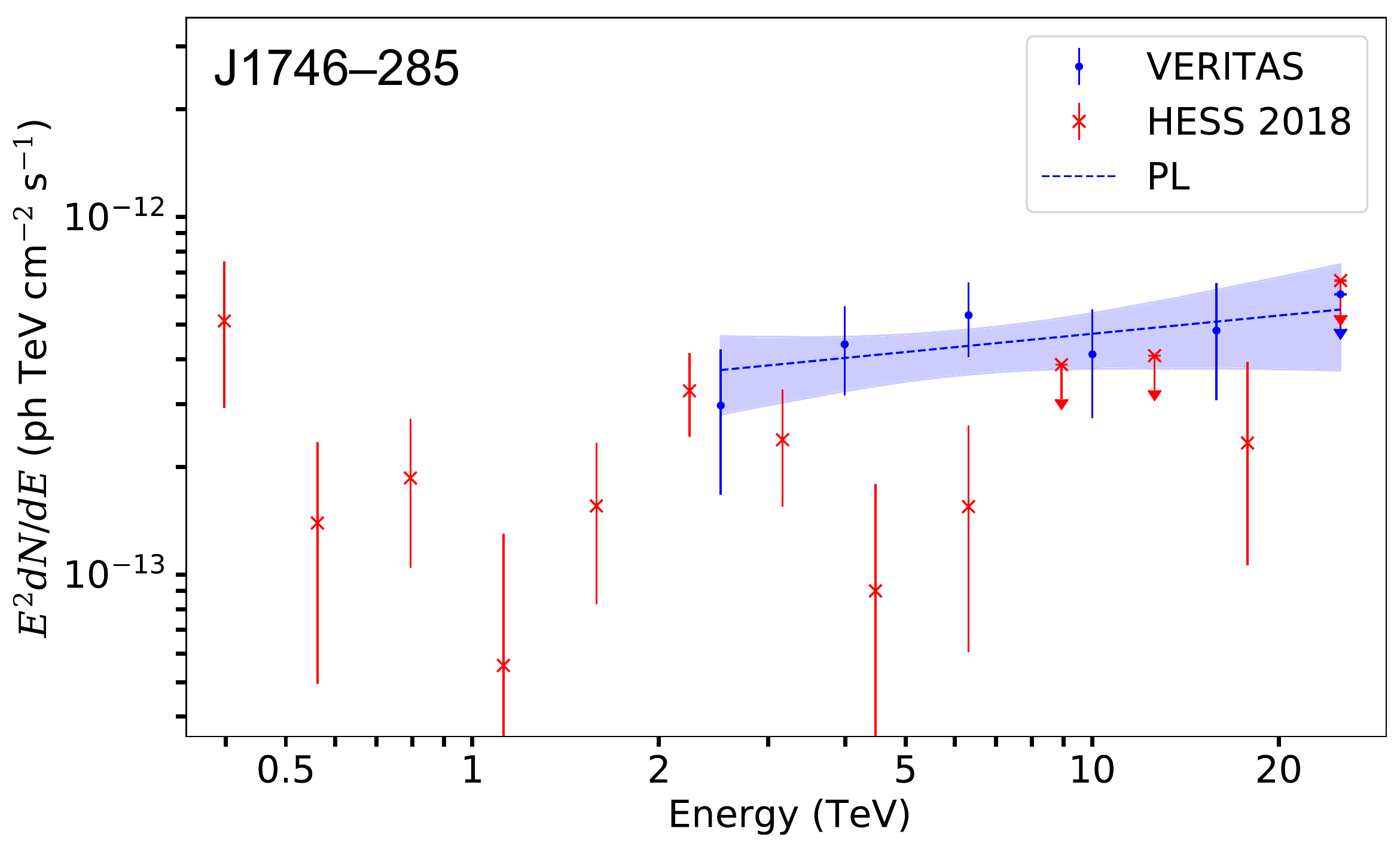}
  \caption{Differential energy spectrum at the position of HESS J1746--285 \citep{abdalla2018b}, as measured by VERITAS (blue), with the H.E.S.S. spectrum shown in red \citep{abdalla2018a}.
  Error bars represent $1\sigma$ uncertainties in the flux.
  The downward arrow represents a 95\% upper limit.
  The blue dashed line is the best-fit power law with index $-1.83 \pm 0.22$
  The shaded region represents the $1\sigma$ confidence band on the model fit.}
  \label{fig:j1746-289_spectrum}
\end{SCfigure}
VERITAS has previously reported the detection of VER J1746-289 \citep{archer2016}, adjacent to the GC.
HESS and MAGIC have also detected sources in this vicinity, named HESS J1746-285 and MAGIC J1746.4-2853, respectively \citep{abdalla2018b,ahnen2017}.
We consider the HESS position (with which the MAGIC position is in agreement) to be more accurate than our measurement in \citep{archer2016}, due to their simultaneous modeling of point sources and multiple diffuse emission components.
We therefore perform our spectral analysis using the position of HESS J1746--285 \citep{abdalla2018b}.
Its spectrum can be described by a power law with $N_0=(1.51\pm 0.22)\times10^{-14}\dflux$, $\Gamma=-1.83\pm0.22$, and $E_0=5.3$ TeV, in the energy range 2--20 TeV.
We caution that contamination from J1745-290 and the diffuse ridge emission could contribute up to $\sim$50\% of the total integrated flux in the measured spectrum.

Potential associations of this source at different wavelengths include the \textit{Fermi}-LAT source 4FGL J1746.4-2852, the Galactic radio arc, and PWN candidate G0.13-0.11, detected by \textit{Chandra}.

\section{Conclusions}
The development of new methods has greatly improved the accuracy of LZA analyses by VERITAS.
Recently, many of these methods have been successfully applied for the first time to the complicated GC region.
In particular, we make the first VERITAS measurement of the diffuse ridge emission.
Future work will include more flexible, likelihood-based methods for handling extended sources, e.g. to be used in searches for dark matter in the GC region.

TeV observations of the sources in the GC remain important for numerous areas in astrophysics.
For this reason, the GC constitutes one of the Key Science Projects for the Cherenkov Telescope Array, which will observe the region with unprecedented spatial and spectral resolutions. 

\section{Acknowledgements}
This research is supported by grants from the U.S. Department of Energy Office of Science, the U.S. National Science Foundation and the Smithsonian Institution, by NSERC in Canada, and by the Helmholtz Association in Germany. This research used resources provided by the Open Science Grid, which is supported by the National Science Foundation and the U.S. Department of Energy's Office of Science, and resources of the National Energy Research Scientific Computing Center (NERSC), a U.S. Department of Energy Office of Science User Facility operated under Contract No. DE-AC02-05CH11231. We acknowledge the excellent work of the technical support staff at the Fred Lawrence Whipple Observatory and at the collaborating institutions in the construction and operation of the instrument.

A portion of our support came from these awards from the National Science Foundation:
PHY-1307171, ``Particle Astrophysics with VERITAS and Defining Scientific Horizons for CTA"
and
PHY-1607491, ``Particle Astrophysics with VERITAS and Development for CTA."

\bibliographystyle{apj}
\bibliography{icrc}

\clearpage
\section*{Full Authors List: \Coll\ Collaboration}

\scriptsize
\noindent
C.~B.~Adams$^{1}$,
A.~Archer$^{2}$,
W.~Benbow$^{3}$,
A.~Brill$^{1}$,
J.~H.~Buckley$^{4}$,
M.~Capasso$^{5}$,
J.~L.~Christiansen$^{6}$,
A.~J.~Chromey$^{7}$, 
M.~Errando$^{4}$,
A.~Falcone$^{8}$,
K.~A.~Farrell$^{9}$,
Q.~Feng$^{5}$,
G.~M.~Foote$^{10}$,
L.~Fortson$^{11}$,
A.~Furniss$^{12}$,
A.~Gent$^{13}$,
G.~H.~Gillanders$^{14}$,
C.~Giuri$^{15}$,
O.~Gueta$^{15}$,
D.~Hanna$^{16}$,
O.~Hervet$^{17}$,
J.~Holder$^{10}$,
B.~Hona$^{18}$,
T.~B.~Humensky$^{1}$,
W.~Jin$^{19}$,
P.~Kaaret$^{20}$,
M.~Kertzman$^{2}$,
D.~Kieda$^{18}$,
T.~K.~Kleiner$^{15}$,
S.~Kumar$^{16}$,
M.~J.~Lang$^{14}$,
M.~Lundy$^{16}$,
G.~Maier$^{15}$,
C.~E~McGrath$^{9}$,
P.~Moriarty$^{14}$,
R.~Mukherjee$^{5}$,
D.~Nieto$^{21}$,
M.~Nievas-Rosillo$^{15}$,
S.~O'Brien$^{16}$,
R.~A.~Ong$^{22}$,
A.~N.~Otte$^{13}$,
S.~R. Patel$^{15}$,
S.~Patel$^{20}$,
K.~Pfrang$^{15}$,
M.~Pohl$^{23,15}$,
R.~R.~Prado$^{15}$,
E.~Pueschel$^{15}$,
J.~Quinn$^{9}$,
K.~Ragan$^{16}$,
P.~T.~Reynolds$^{24}$,
D.~Ribeiro$^{1}$,
E.~Roache$^{3}$,
J.~L.~Ryan$^{22}$,
I.~Sadeh$^{15}$,
M.~Santander$^{19}$,
G.~H.~Sembroski$^{25}$,
R.~Shang$^{22}$,
D.~Tak$^{15}$,
V.~V.~Vassiliev$^{22}$,
A.~Weinstein$^{7}$,
D.~A.~Williams$^{17}$,
and 
T.~J.~Williamson$^{10}$\\
\noindent
$^{1}${Physics Department, Columbia University, New York, NY 10027, USA}
$^{2}${Department of Physics and Astronomy, DePauw University, Greencastle, IN 46135-0037, USA}
$^{3}${Center for Astrophysics $|$ Harvard \& Smithsonian, Cambridge, MA 02138, USA}
$^{4}${Department of Physics, Washington University, St. Louis, MO 63130, USA}
$^{5}${Department of Physics and Astronomy, Barnard College, Columbia University, NY 10027, USA}
$^{6}${Physics Department, California Polytechnic State University, San Luis Obispo, CA 94307, USA} 
$^{7}${Department of Physics and Astronomy, Iowa State University, Ames, IA 50011, USA}
$^{8}${Department of Astronomy and Astrophysics, 525 Davey Lab, Pennsylvania State University, University Park, PA 16802, USA}
$^{9}${School of Physics, University College Dublin, Belfield, Dublin 4, Ireland}
$^{10}${Department of Physics and Astronomy and the Bartol Research Institute, University of Delaware, Newark, DE 19716, USA}
$^{11}${School of Physics and Astronomy, University of Minnesota, Minneapolis, MN 55455, USA}
$^{12}${Department of Physics, California State University - East Bay, Hayward, CA 94542, USA}
$^{13}${School of Physics and Center for Relativistic Astrophysics, Georgia Institute of Technology, 837 State Street NW, Atlanta, GA 30332-0430}
$^{14}${School of Physics, National University of Ireland Galway, University Road, Galway, Ireland}
$^{15}${DESY, Platanenallee 6, 15738 Zeuthen, Germany}
$^{16}${Physics Department, McGill University, Montreal, QC H3A 2T8, Canada}
$^{17}${Santa Cruz Institute for Particle Physics and Department of Physics, University of California, Santa Cruz, CA 95064, USA}
$^{18}${Department of Physics and Astronomy, University of Utah, Salt Lake City, UT 84112, USA}
$^{19}${Department of Physics and Astronomy, University of Alabama, Tuscaloosa, AL 35487, USA}
$^{20}${Department of Physics and Astronomy, University of Iowa, Van Allen Hall, Iowa City, IA 52242, USA}
$^{21}${Institute of Particle and Cosmos Physics, Universidad Complutense de Madrid, 28040 Madrid, Spain}
$^{22}${Department of Physics and Astronomy, University of California, Los Angeles, CA 90095, USA}
$^{23}${Institute of Physics and Astronomy, University of Potsdam, 14476 Potsdam-Golm, Germany}
$^{24}${Department of Physical Sciences, Munster Technological University, Bishopstown, Cork, T12 P928, Ireland}
$^{25}${Department of Physics and Astronomy, Purdue University, West Lafayette, IN 47907, USA}

\end{document}